\documentclass[dvips]{article}
\usepackage{lscape,epsfig}
\usepackage{amssymb}
\usepackage{amsmath}
\usepackage{rotating}
\DeclareSymbolFont{ppa}{OT1}{ppl}{m}{it}
\DeclareMathSymbol{\vv}{\mathalpha}{ppa}{'166}

\begin{document}

\newcommand{\dd}{\,{\rm d}}
\newcommand{\ie}{{\it i.e.},\,}
\newcommand{\etal}{{\it et al.\ }}
\newcommand{\eg}{{\it e.g.},\,}
\newcommand{\cf}{{\it cf.\ }}
\newcommand{\vs}{{\it vs.\ }}
\newcommand{\zdot}{\makebox[0pt][l]{.}}
\newcommand{\up}[1]{\ifmmode^{\rm #1}\else$^{\rm #1}$\fi}
\newcommand{\dn}[1]{\ifmmode_{\rm #1}\else$_{\rm #1}$\fi}
\newcommand{\upd}{\up{d}}
\newcommand{\uph}{\up{h}}
\newcommand{\upm}{\up{m}}
\newcommand{\ups}{\up{s}}
\newcommand{\arcd}{\ifmmode^{\circ}\else$^{\circ}$\fi}
\newcommand{\arcm}{\ifmmode{'}\else$'$\fi}
\newcommand{\arcs}{\ifmmode{''}\else$''$\fi}
\newcommand{\MS}{{\rm M}\ifmmode_{\odot}\else$_{\odot}$\fi}
\newcommand{\RS}{{\rm R}\ifmmode_{\odot}\else$_{\odot}$\fi}
\newcommand{\LS}{{\rm L}\ifmmode_{\odot}\else$_{\odot}$\fi}

\newcommand{\Abstract}[2]{{\footnotesize\begin{center}ABSTRACT\end{center}
\vspace{1mm}\par#1\par
\noindent
{~}{\it #2}}}

\newcommand{\TabCap}[2]{\begin{center}\parbox[t]{#1}{\begin{center}
  \small {\spaceskip 2pt plus 1pt minus 1pt T a b l e}
  \refstepcounter{table}\thetable \\[2mm]
  \footnotesize #2 \end{center}}\end{center}}

\newcommand{\TableSep}[2]{\begin{table}[p]\vspace{#1}
\TabCap{#2}\end{table}}

\newcommand{\FigCap}[1]{\footnotesize\par\noindent Fig.\  %
  \refstepcounter{figure}\thefigure. #1\par}

\newcommand{\TableFont}{\footnotesize}
\newcommand{\TableFontIt}{\ttit}
\newcommand{\SetTableFont}[1]{\renewcommand{\TableFont}{#1}}

\newcommand{\MakeTable}[4]{\begin{table}[htb]\TabCap{#2}{#3}
  \begin{center} \TableFont \begin{tabular}{#1} #4
  \end{tabular}\end{center}\end{table}}

\newcommand{\MakeTableSep}[4]{\begin{table}[p]\TabCap{#2}{#3}
  \begin{center} \TableFont \begin{tabular}{#1} #4
  \end{tabular}\end{center}\end{table}}

\newenvironment{references}%
{
\footnotesize \frenchspacing
\renewcommand{\thesection}{}
\renewcommand{\in}{{\rm in }}
\renewcommand{\AA}{Astron.\ Astrophys.}
\newcommand{\AAS}{Astron.~Astrophys.~Suppl.~Ser.}
\newcommand{\ApJ}{Astrophys.\ J.}
\newcommand{\ApJS}{Astrophys.\ J.~Suppl.~Ser.}
\newcommand{\ApJL}{Astrophys.\ J.~Letters}
\newcommand{\AJ}{Astron.\ J.}
\newcommand{\IBVS}{IBVS}
\newcommand{\PASP}{P.A.S.P.}
\newcommand{\Acta}{Acta Astron.}
\newcommand{\MNRAS}{MNRAS}
\renewcommand{\and}{{\rm and }}
\section{{\rm REFERENCES}}
\sloppy \hyphenpenalty10000
\begin{list}{}{\leftmargin1cm\listparindent-1cm
\itemindent\listparindent\parsep0pt\itemsep0pt}}%
{\end{list}\vspace{2mm}}

\def\TYLDA{~}
\newlength{\DW}
\settowidth{\DW}{0}
\newcommand{\dw}{\hspace{\DW}}

\newcommand{\refitem}[5]{\item[]{#1} #2%
\def\REFARG{#3}\ifx\REFARG\TYLDA\else, {\it#3}\fi
\def\REFARG{#4}\ifx\REFARG\TYLDA\else, {\bf#4}\fi
\def\REFARG{#5}\ifx\REFARG\TYLDA\else, {#5}\fi.}

\newcommand{\Section}[1]{\section{#1}}
\newcommand{\Subsection}[1]{\subsection{#1}}
\newcommand{\Acknow}[1]{\par\vspace{5mm}{\bf Acknowledgments.} #1}
\pagestyle{myheadings}

\newfont{\bb}{ptmbi8t at 12pt}
\newcommand{\xrule}{\rule{0pt}{2.5ex}}
\newcommand{\xxrule}{\rule[-1.8ex]{0pt}{4.5ex}}
\def\thefootnote{\fnsymbol{footnote}}

\begin{center}
{\Large\bf Searching for Potential Mergers among 22~500 Eclipsing Binary
Stars in the OGLE-III Galactic Bulge Fields\footnote{Based on observations
obtained with the 1.3-m Warsaw telescope at the Las Campanas Observatory of the
Carnegie Institution for Science.}}
\vskip1cm
{\bf
P.~~P~i~e~t~r~u~k~o~w~i~c~z$^1$,~~I.~~S~o~s~z~y~\'n~s~k~i$^1$,~~A.~~U~d~a~l~s~k~i$^1$,\\
~~M.~K.~~S~z~y~m~a~\'n~s~k~i$^1$,~~{\L}.~~W~y~r~z~y~k~o~w~s~k~i$^{1}$,~~R.~~P~o~l~e~s~k~i$^{1,2}$,\\
~~S.~~K~o~z~{\l}~o~w~s~k~i$^1$,~~J.~~S~k~o~w~r~o~n$^1$,~~P.~~M~r~\'o~z$^1$,~~M.~P~a~w~l~a~k$^1$,\\
and~~K.~~U~l~a~c~z~y~k$^{1,3}$\\}
\vskip3mm
{
$^1$ Warsaw University Observatory, Al. Ujazdowskie 4, 00-478 Warszawa, Poland\\
e-mail: pietruk@astrouw.edu.pl\\
$^2$ Department of Astronomy, Ohio State University, 140 W. 18th Ave.,\\
Columbus, OH 43210, USA\\
$^3$ Department of Physics, University of Warwick, Coventry CV4 7AL, UK\\
}
\end{center}

\Abstract{Inspired by the discovery of the red nova V1309 Sco
(Nova Scorpii 2008) and the fact that its progenitor was a binary system
with a rapidly decreasing orbital period, we have searched for
period changes in OGLE binary stars. We have selected a sample of 22~462
short-period ($P_{\rm orb}<4$~d) eclipsing binary stars observed
toward the Galactic bulge by the OGLE-III survey in years 2001--2009.
This dataset was extended with photometry from OGLE-II (1997--2000)
and the first six years of OGLE-IV (2010--2015). For some stars,
the data were supplemented with OGLE-I photometry (1992--1995).
After close inspection of the whole sample we have found 56 systems with
realistic period decrease and 52 systems with realistic period increase.
We have also recognized 35 systems with cyclic period variations.
The highest negative period change rate of $-1.943\times 10^{-4}$~d/y
has been detected in detached eclipsing binary OGLE-BLG-ECL-139622 with
$P_{\rm orb}=2.817$~d, while all other found systems are contact
binaries with orbital periods mostly shorter than 1.0~d. For 22 our systems
with decreasing orbital period the absolute rate is higher than the value
reported recently for eclipsing binary KIC 9832227. Interestingly,
there is an excess of systems with high negative period change rate over
systems with positive rate. We cannot exclude the possibility that some
of the contact binaries with relatively long orbital period and high
negative period change rate will merge in the future. However, our results
rather point to the presence of tertiary companions in the observed
systems and/or spot activity on the surface of the binary components.}

{Galaxy: bulge -- Galaxy: disk -- binaries: eclipsing}


\Section{Introduction}

Red novae belong to a very intriguing class of rarely observed luminous
transients. They are associated with a dynamical phase of common envelope
evolution (Soker and Tylenda 2003), albeit alternative explanations have
been also proposed, such as classical nova eruption with a slowly moving,
massive envelope (Shara \etal 2010), or explosion of the star in the
dust-enshrouded phase of the evolution at the extremum of the asymptotic
giant branch (Thompson \etal 2009), or intensive mass loss from
the binary system through the outer Lagrangian point (Pejcha 2014).
So far, only a handful number of this kind of transients have been noted
in the Milky Way: V4332 Sgr (Hayashi \etal 1994), V838 Mon (Brown \etal 2002),
V1309 Sco (Nakano \etal 2008), and OGLE-2002-BLG-360 (Tylenda \etal 2013).
Thanks to precise high-cadence OGLE photometry covering seven years before
the eruption of the red nova V1309 Sco, it was realized that the progenitor
was a contact eclipsing binary with the orbital period shrinking
from about 1.438~d in 2002 to 1.425~d in 2007 (Tylenda \etal 2011).
Before the eruption, the light curve shape of the object transformed
from a double wave to single wave which indicated the immersion of two
stellar bodies in an elongated common envelope. According to the
estimation made by Kochanek \etal (2014), the number of Galactic events
like the V1309 Sco outburst is about once per decade.

The main goal of our work is searching for candidates for future
stellar mergers or systems with high negative period change rate
among eclipsing binaries observed by the OGLE survey toward
the Milky Way bulge. From the sample of eclipsing binaries,
we also select systems with positive period change rate and
systems with noticeable cyclic period variations.


\Section{Observations}

The Optical Gravitational Lensing Experiment (OGLE) is a long-term
wide-field variability sky survey launched in 1992 with the
original aim of searching for microlensing events (Udalski \etal 1992).
The survey is conducted at Las Campanas Observatory which is
operated by the Carnegie Institution for Science. OGLE uses
the 1.3-m Warsaw Telescope. Since 2010 the project
is in its fourth phase, OGLE-IV (Udalski \etal 2015), collecting
data with a 32-CCD mosaic camera of a field of view of 1.4 deg$^2$.
Currently, OGLE monitors over 3000~deg$^2$ of the sky.
Previous phases of the project were conducted in the following years:
1992--1995 (OGLE-I), 1997--2000 (OGLE-II), and 2001--2009 (OGLE-III).
In 2016 the OGLE database exceeded $10^{12}$ single photometric measurements.
The survey has discovered and classified nearly one million genuine
variable stars toward the Galactic bulge, Galactic disk, and Magellanic
Clouds (\eg Soszy\'nski \etal 2013,2014,2015,2016, Pietrukowicz \etal 2013,
Mr\'oz \etal 2015).

Eclipsing binary systems, among which we look for potential mergers,
were observed by OGLE-III in the direction of the Galactic bulge.
In the third phase of OGLE, an eight-CCD mosaic camera with a field
of view of about 0.35 deg$^2$ was attached to the Warsaw Telescope
(Udalski \etal 2003). Angular pixel size in OGLE-III and OGLE-IV
is the same: 0\zdot\arcs26. During the OGLE-III phase 267 fields
covering about 92~deg$^2$ toward the Milky Way bulge were observed
(Szyma\'nski \etal 2011). In OGLE-II, a single CCD camera with a pixel
size of 0\zdot\arcs42 was used in driftscan mode (Udalski \etal 1997).
In that phase, 49 Galactic bulge fields of $14\zdot\arcm2\times57\arcm$
each covering a total area of around 11~deg$^2$ were monitored.
The bulge coverage of OGLE-I, conducted on the 1.0-m Swope telescope
also at Las Campanas Observatory, was even much smaller: 18 fields
of $15\arcm\times15\arcm$ each covered about 1.1~deg$^2$.
OGLE monitors the sky mainly in the $I$-band, while $V$-band observations
are collected to secure color information of the objects and for accurate
transformation to the standard Johnson-Cousins system. The number of
$I$($V$)-band measurements in the most frequently observed Galactic
bulge fields is the following: 261(44) in OGLE-I, 568(16) in OGLE-II,
2540(34) in OGLE-III, and 12~889(144) in OGLE-IV (2010--2015).
Reduction of the OGLE data is performed with the difference image
analysis (DIA) technique (Alard and Lupton 1998, Wo\'zniak 2000).


\Section{Binary Systems Selection}

For the purpose of our analysis we selected 53 OGLE-III Galactic
bulge fields observed in the $I$-band for at least 6 seasons
with a minimum number of 40 epochs per season. Location of these
fields in Galactic coordinates is presented in Fig.~1.
Over 28 million detections in the brightness range $12.6<I<19.0$~mag
were a subject of the initial period search with the FNPEAKS
code\footnote{http://helas.astro.uni.wroc.pl/deliverables.php?lang=en\&active=fnpeaks}
for each season separately. For further analysis we left about
14 million detections with assessed periods in all well-covered seasons.
From this sample we removed detections with signals around 1/3~d,
1/2~d, and 1~d being very likely daily aliases. Since we concentrate
on short-period systems only, we removed objects with the initially
detected periods longer than 2~d or orbital periods $P_{\rm orb}>4$~d.
After some verification tests, we decided to work further on 1\% of stars
with the highest variability signal. The initial periods for about 137~500
stars were corrected with the TATRY code (Schwarzenberg-Czerny 1996).
Cross-matching of this sample with the list of OGLE-III Galactic bulge
RR Lyr-type stars (Soszy\'nski \etal 2011) led to the rejection
of about 2\% of stars. RR Lyr stars, particularly of RRc type showing
close-to-sinusoidal light curve shapes and with periods of a few tens
of day, could contaminate our sample. We made a visual inspection
of $I$-band light curves of around 134~700 detections. This time-consuming
operation allowed us to reject other contaminants from the sample, such as
spotted variables, rotating variables, and $\delta$~Sct-type pulsators.
Importantly, the inspection helped us to verify the orbital periods
of candidates for eclipsing binaries. We corrected our list of candidate
OGLE-III eclipsing variables for artifacts and duplicates detected
in adjacent fields.

In the next step, we searched for OGLE-II and OGLE-IV counterparts.
OGLE-IV data used cover six seasons, from 2010 to 2015.
About 43\% and 95\% of OGLE-III bulge binaries are present in the
OGLE-II and OGLE-IV images, respectively. As in the case of the
OGLE-III data, we took into account only binaries with at least 40
$I$-band measurements per season. Several times more frequent OGLE-IV
observations allowed us to verify the classification of the variables
and to correct the orbital periods determined from the OGLE-III data.

The final number of detected eclipsing binaries is 22~462.
For 19~885 binaries, the OGLE-III data are extended with photometry from
OGLE-IV. For 8788 binaries, the data are also extended with photometry from
OGLE-II. For 7825 eclipsing binaries our dataset covers 20 years
of OGLE-II, OGLE-III, and OGLE-IV. In the case of the most interesting
systems, this dataset was supplemented with OGLE-I photometry increasing
the time coverage up to 24 years. In our final sample, there are 1657
binaries with solely OGLE-III photometry. All light curves were
cleaned from outlying points by phasing and binning the data. After some
tests we set the cleaning limit at a mild level of $5\sigma$ to avoid
rejection of good data points in the case of systems with possibly
high period change rates. For objects that are present in the OGLE
collection of eclipsing binary stars toward the Galactic bulge published
by Soszy\'nski \etal (2016), we use their format: OGLE-BLG-ECL-NNNNNN.
For objects that are absent in that collection, we left the standard
format used in the OGLE database: FIELD.CHIP.ID.

\begin{figure}[htb]
\centerline{\includegraphics[angle=0,width=130mm]{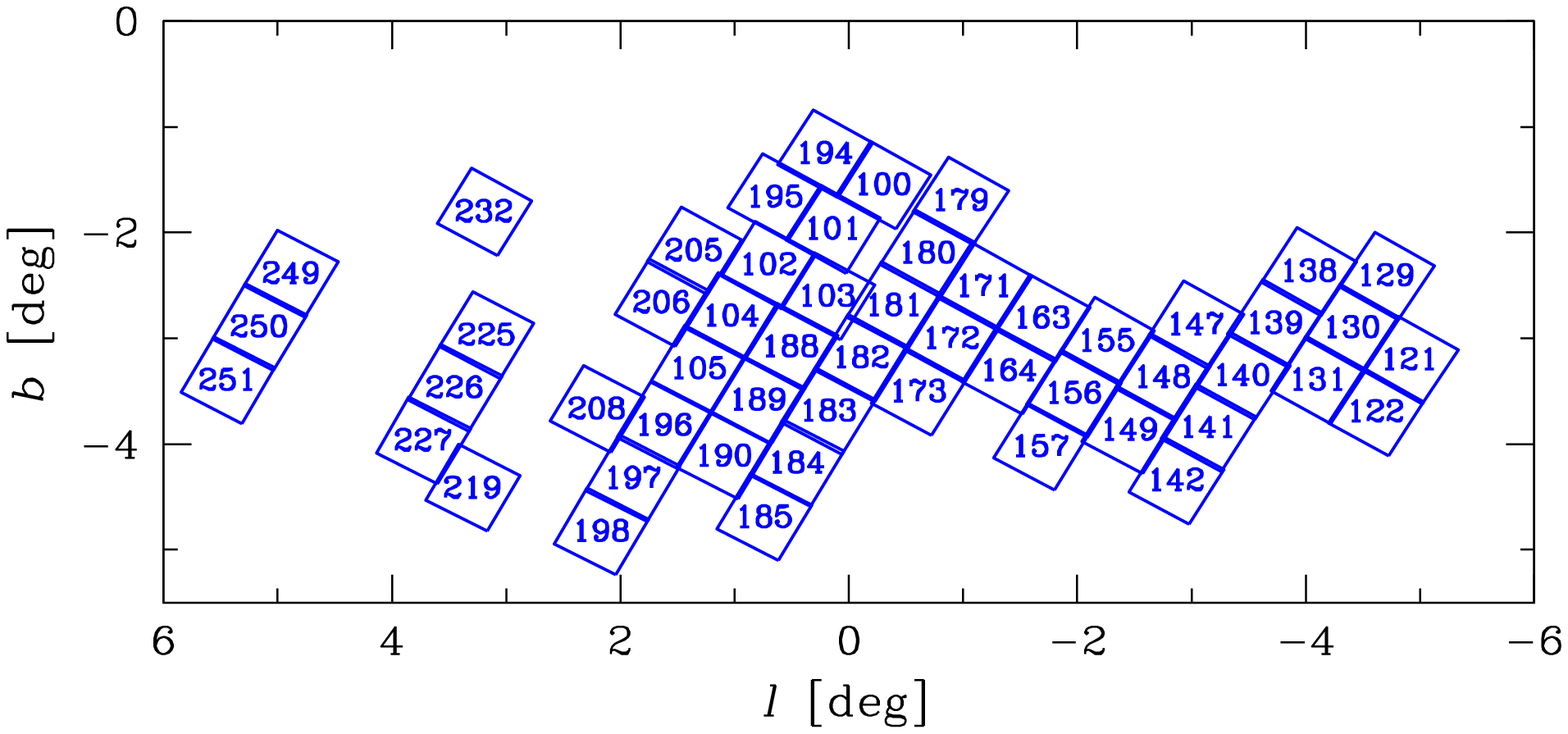}}
\FigCap{Location of 53 OGLE-III Galactic bulge fields selected
for eclipsing binary stars searches.}
\end{figure}


\Section{Period Changes}

A simple method was applied to find systems with reliable period
change rates and candidates for possible mergers in the sample
of 22~462 eclipsing binaries. We searched for negative as well as
positive linear period changes using $I$-band photometry from
OGLE-II, OGLE-III, and OGLE-IV, depending on the coverage.
OGLE-III and OGLE-IV data were divided into seasons.
After some tests we decided not to divide the less rich OGLE-II data.
For each OGLE-III/IV season and the whole four-year OGLE-II dataset
accurate orbital periods were determined using the TATRY code
(Schwarzenberg-Czerny 1996). Moments corresponding to each
period of time were calculated as average from the epochs.
Our approach allowed us to avoid possible problems with variations
in mean brightness, amplitude, and light curve shapes (due to star
spots), also problems with the presence of some remnant outlying
points near eclipses and small magnitude offsets between the photometry
from different OGLE phases to all of which the classical $O-C$ method
is sensitive (\eg Conroy \etal 2014, Gies \etal 2015 in the application
to data from the Kepler satellite). After fitting a linear regression
to all obtained ``period change curves'', we made a close inspection
of 2403 systems for which $|\dot P_{\rm orb}|>5\sigma_{\dot P{\rm orb}}$.
In many period change curves with particularly high rate derived from the
automated fit, we noticed outlying points corresponding to less frequently
observed seasons. After the inspection we found only 59 systems with
realistic period decrease and 53 systems with realistic period increase.
We identified 35 systems with evident cyclic period variations.

In the last stage, we verified whether some of the found interesting
binaries had been observed by OGLE-I in years 1992--1995. Photometry
was collected for twelve of these binaries. It turned out that three
candidate systems with decreasing period and one candidate system
with increasing period seem to show rather non-monotonic variations.
The final number of detected systems with the increasing, decreasing,
and cyclic period changes is 56, 52, and 35 objects, respectively.

We recognized four of our systems in the list of 569 contact binaries with
reliable period change rates determined by Kubiak \etal (2006) based on
OGLE data from years 1992--2005. Systems BW1.125206 = OGLE-BLG-ECL-265310
and BWC.169286 = OGLE-BLG-ECL-276943 were found, at that time, to have
the decreasing period, while systems BW7.159932 = OGLE-BLG-ECL-288099
and BW4.5243 = OGLE-BLG-ECL-279991 to have the increasing period.
However, observations spanning 24 years indicate non-monotonic
variations of the period in these four binaries.

Period change curves together with phased $I$-band light curves for seven
of our binary stars with the highest negative orbital period change rate are
presented in Fig.~2. In Table~1, we list basic parameters of all 56 systems
with reliable negative period changes sorted from the highest absolute rate.
The given orbital periods correspond roughly to the middle of the OGLE-III
phase (year 2005). System OGLE-BLG-ECL-139622 with the highest derived negative
period change rate of $\dot P_{\rm orb}=-1.943\times10^{-4}$~d/y is a detached
eclipsing binary of Algol (EA) type. Unfortunately, this object was monitored
only in the OGLE-III phase. New time-series data would help in verification
of the observed trend. Other systems with the decreasing orbital period are
contact binaries. All of them have absolute rates at least one order
of magnitude lower than OGLE-BLG-ECL-139622. Interestingly, high negative
period change rates are observed in contact systems with relatively
long orbital periods: all four contact binaries with $P_{\rm orb}>1.0$~d
have $|\dot P_{\rm orb}|>10^{-5}$~d/y; all seven contact binaries with
$P_{\rm orb}>0.8$~d have $|\dot P_{\rm orb}|>6\times10^{-6}$~d/y.

In Fig.~3, we present seven binaries with the highest derived positive
period change rate. Table~2 provides information on all 52 systems with the
increasing orbital period. All such systems have the orbital period
$<0.8$~d and the period change rate $<10^{-5}$~d/y or lower
than the absolute value in eight systems with the most rapid negative
period changes. Distributions of systems with negative and positive period
change rates are compared in Fig.~4. All detected systems with the increasing
orbital period are contact binaries.

Finally, in Fig.~5 and Table~3, we present systems with cyclic period
variations. We assess the cycle period for each of the system.
All these systems are contact binaries with $P_{\rm orb}<0.9$~d.

\begin{figure}[htb]
\centerline{\includegraphics[angle=0,width=130mm]{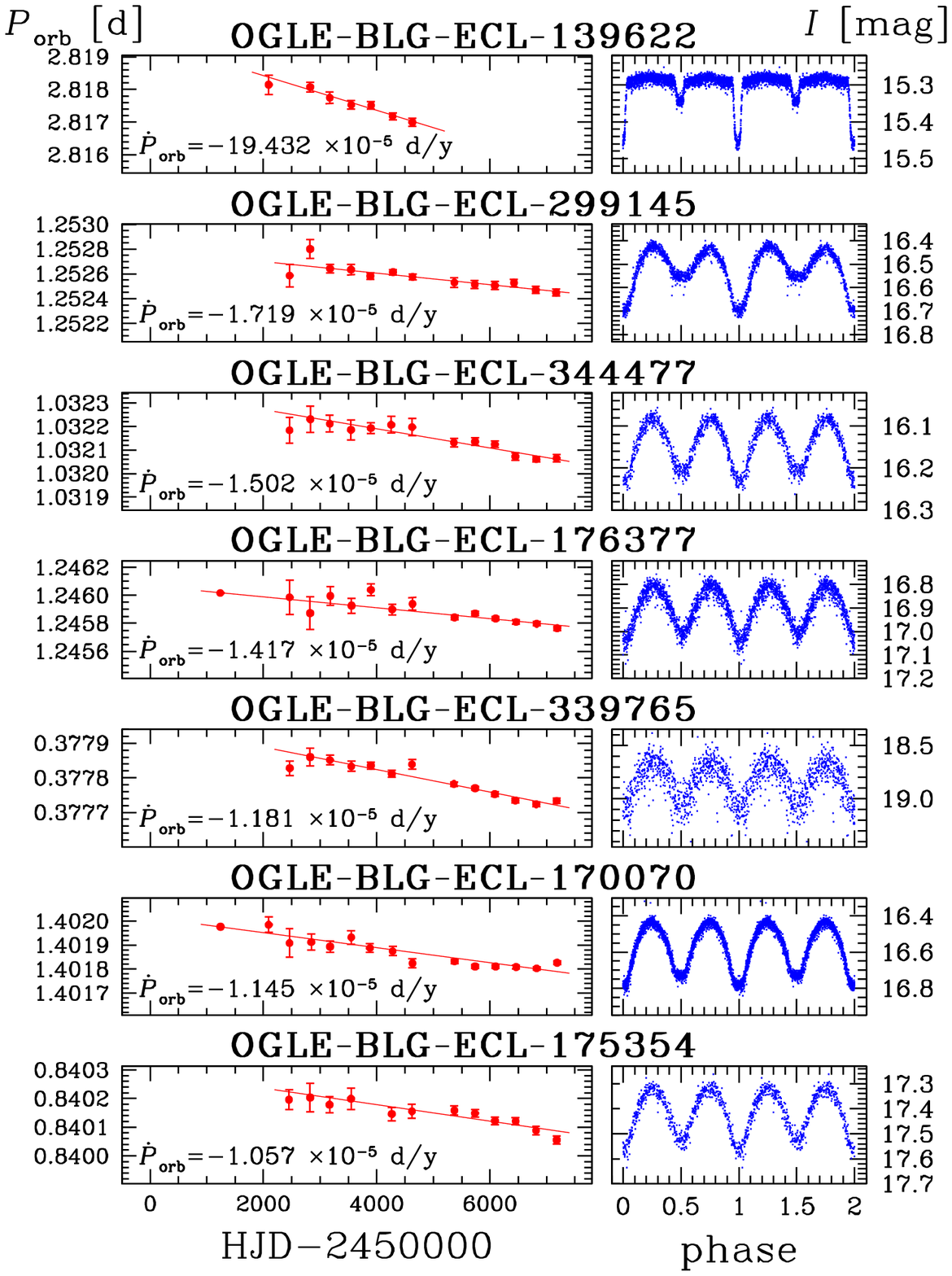}}
\FigCap{Binary systems with the highest negative period change rates.
Presented OGLE-III light curves are phased with constant period
determined for this dataset.}
\end{figure}

\begin{table}[h!]
\centering
\caption{\small Parameters of the systems with derived negative period change rate}
\medskip
{\footnotesize
\begin{tabular}{ccccccc}
\hline
OGLE-BLG-ECL- &   RA    &    Dec  & $I_{\rm max}$ & $V-I$ & $P_{\rm orb}$ & $\dot P_{\rm orb}$ \\
              & J2000.0 & J2000.0 &     [mag]     & [mag] &      [d]      &         [d/y] \\
\hline
139622 & $17\uph51\upm33\zdot\ups33$ & $-29\arcd54\arcm38\zdot\arcs7$ & 15.28 & 1.41 & 2.817436 & $-1.94\pm0.28$~E$-4$ \\
299145 & $18\uph05\upm34\zdot\ups20$ & $-29\arcd40\arcm21\zdot\arcs5$ & 16.42 & 1.26 & 1.252603 & $-1.72\pm0.24$~E$-5$ \\
344477 & $18\uph10\upm24\zdot\ups89$ & $-28\arcd34\arcm26\zdot\arcs5$ & 16.08 & 1.23 & 1.032195 & $-1.50\pm0.18$~E$-5$ \\
176377 & $17\uph54\upm41\zdot\ups59$ & $-29\arcd29\arcm18\zdot\arcs1$ & 16.80 & 1.74 & 1.245946 & $-1.42\pm0.05$~E$-5$ \\
339765 & $18\uph09\upm52\zdot\ups64$ & $-27\arcd25\arcm53\zdot\arcs1$ & 18.62 & 1.47 & 0.377840 & $-1.18\pm0.08$~E$-5$ \\
170070 & $17\uph54\upm09\zdot\ups38$ & $-29\arcd42\arcm53\zdot\arcs7$ & 16.43 & 1.85 & 1.401907 & $-1.15\pm0.03$~E$-5$ \\
175354 & $17\uph54\upm36\zdot\ups55$ & $-33\arcd44\arcm51\zdot\arcs9$ & 17.31 & 1.58 & 0.840163 & $-1.06\pm0.16$~E$-5$ \\
154749 & $17\uph52\upm54\zdot\ups10$ & $-29\arcd08\arcm29\zdot\arcs7$ & 17.55 & 2.45 & 0.932247 & $-1.04\pm0.18$~E$-5$ \\
220028 & $17\uph58\upm24\zdot\ups67$ & $-29\arcd57\arcm42\zdot\arcs0$ & 16.68 & 1.53 & 0.920792 & $-6.02\pm1.10$~E$-6$ \\
172659 & $17\uph54\upm22\zdot\ups13$ & $-30\arcd01\arcm41\zdot\arcs9$ & 18.10 & 1.77 & 0.626226 & $-5.85\pm0.21$~E$-6$ \\
195664 & $17\uph56\upm19\zdot\ups97$ & $-30\arcd55\arcm40\zdot\arcs9$ & 18.55 & 1.71 & 0.436900 & $-4.56\pm0.19$~E$-6$ \\
192939 & $17\uph56\upm06\zdot\ups40$ & $-29\arcd29\arcm21\zdot\arcs4$ & 18.23 & 2.09 & 0.531934 & $-4.40\pm0.20$~E$-6$ \\
243151 & $18\uph00\upm27\zdot\ups01$ & $-29\arcd27\arcm41\zdot\arcs1$ & 18.41 & 1.62 & 0.403628 & $-4.35\pm0.50$~E$-6$ \\
347999 & $18\uph10\upm49\zdot\ups69$ & $-29\arcd25\arcm35\zdot\arcs8$ & 18.39 & 1.31 & 0.365383 & $-4.30\pm0.66$~E$-6$ \\
212252 & $17\uph57\upm44\zdot\ups48$ & $-31\arcd03\arcm40\zdot\arcs8$ & 18.71 & 1.95 & 0.413250 & $-3.88\pm0.17$~E$-6$ \\
168049 & $17\uph53\upm59\zdot\ups85$ & $-30\arcd07\arcm37\zdot\arcs4$ & 17.66 & 1.64 & 0.506522 & $-2.71\pm0.18$~E$-6$ \\
320382 & $18\uph07\upm43\zdot\ups53$ & $-25\arcd42\arcm42\zdot\arcs8$ & 17.31 & 1.68 & 0.388373 & $-2.63\pm0.40$~E$-6$ \\
218785 & $17\uph58\upm18\zdot\ups73$ & $-32\arcd12\arcm24\zdot\arcs8$ & 18.04 & 1.82 & 0.353455 & $-2.53\pm0.36$~E$-6$ \\
119460 & $17\uph49\upm11\zdot\ups03$ & $-35\arcd01\arcm02\zdot\arcs8$ & 17.12 & 1.41 & 0.394387 & $-2.49\pm0.36$~E$-6$ \\
351804 & $18\uph11\upm16\zdot\ups73$ & $-25\arcd42\arcm55\zdot\arcs4$ & 16.71 & 1.46 & 0.385420 & $-2.12\pm0.26$~E$-6$ \\
130708 & $17\uph50\upm35\zdot\ups67$ & $-29\arcd19\arcm13\zdot\arcs6$ & 17.49 & 2.10 & 0.366826 & $-2.05\pm0.39$~E$-6$ \\
149279 & $17\uph52\upm26\zdot\ups28$ & $-32\arcd13\arcm12\zdot\arcs7$ & 17.62 & 2.08 & 0.314728 & $-2.03\pm0.32$~E$-6$ \\
183004 & $17\uph55\upm16\zdot\ups99$ & $-31\arcd03\arcm52\zdot\arcs4$ & 15.73 & 1.37 & 0.467429 & $-2.03\pm0.10$~E$-6$ \\
190636 & $17\uph55\upm55\zdot\ups09$ & $-31\arcd00\arcm26\zdot\arcs6$ & 17.63 & 1.85 & 0.305221 & $-1.96\pm0.16$~E$-6$ \\
287671 & $18\uph04\upm31\zdot\ups54$ & $-28\arcd38\arcm06\zdot\arcs5$ & 17.08 & 1.19 & 0.532926 & $-1.94\pm0.07$~E$-6$ \\
201682 & $17\uph56\upm50\zdot\ups80$ & $-29\arcd44\arcm49\zdot\arcs5$ & 17.32 & 1.90 & 0.497314 & $-1.89\pm0.29$~E$-6$ \\
213011 & $17\uph57\upm48\zdot\ups50$ & $-29\arcd21\arcm59\zdot\arcs0$ & 17.65 & 1.94 & 0.573979 & $-1.78\pm0.11$~E$-6$ \\
159116 & $17\uph53\upm16\zdot\ups11$ & $-29\arcd48\arcm59\zdot\arcs7$ & 15.83 & 1.26 & 0.587393 & $-1.72\pm0.06$~E$-6$ \\
229500 & $17\uph59\upm13\zdot\ups78$ & $-30\arcd49\arcm06\zdot\arcs9$ & 16.01 & 1.56 & 0.398365 & $-1.71\pm0.19$~E$-6$ \\
163655 & $17\uph53\upm38\zdot\ups86$ & $-32\arcd55\arcm59\zdot\arcs1$ & 17.00 & 1.70 & 0.492482 & $-1.62\pm0.09$~E$-6$ \\
192437 & $17\uph56\upm03\zdot\ups96$ & $-29\arcd59\arcm12\zdot\arcs6$ & 16.71 & 1.67 & 0.402313 & $-1.60\pm0.07$~E$-6$ \\
233821 & $17\uph59\upm37\zdot\ups95$ & $-28\arcd50\arcm22\zdot\arcs4$ & 15.29 & 1.23 & 0.359920 & $-1.59\pm0.04$~E$-6$ \\
238356 & $18\uph00\upm01\zdot\ups26$ & $-29\arcd09\arcm07\zdot\arcs4$ & 17.23 & 1.52 & 0.292658 & $-1.47\pm0.05$~E$-6$ \\
176119 & $17\uph54\upm40\zdot\ups35$ & $-29\arcd56\arcm26\zdot\arcs1$ & 16.37 & 1.68 & 0.594569 & $-1.44\pm0.06$~E$-6$ \\
181311 & $17\uph55\upm08\zdot\ups47$ & $-29\arcd33\arcm46\zdot\arcs8$ & 16.51 & 1.37 & 0.422617 & $-1.43\pm0.05$~E$-6$ \\
217990 & $17\uph58\upm14\zdot\ups55$ & $-31\arcd30\arcm12\zdot\arcs7$ & 18.39 & 1.83 & 0.412545 & $-1.14\pm0.17$~E$-6$ \\
182438 & $17\uph55\upm14\zdot\ups05$ & $-30\arcd10\arcm56\zdot\arcs0$ & 17.54 & 1.75 & 0.549392 & $-1.11\pm0.09$~E$-6$ \\
280295 & $18\uph03\upm51\zdot\ups13$ & $-27\arcd54\arcm12\zdot\arcs3$ & 16.19 & 1.35 & 0.484237 & $-1.10\pm0.10$~E$-6$ \\
181083 & $17\uph55\upm07\zdot\ups21$ & $-29\arcd58\arcm17\zdot\arcs8$ & 16.20 & 1.45 & 0.349391 & $-1.02\pm0.04$~E$-6$ \\
114280 & $17\uph48\upm28\zdot\ups67$ & $-35\arcd05\arcm09\zdot\arcs9$ & 16.30 & 1.28 & 0.453031 & $-9.99\pm1.64$~E$-7$ \\
106230 & $17\uph47\upm21\zdot\ups97$ & $-34\arcd40\arcm49\zdot\arcs3$ & 16.06 & 1.32 & 0.489019 & $-8.88\pm0.45$~E$-7$ \\
184767 & $17\uph55\upm25\zdot\ups22$ & $-29\arcd56\arcm15\zdot\arcs9$ & 17.38 & 2.42 & 0.473305 & $-8.57\pm1.18$~E$-7$ \\
225278 & $17\uph58\upm52\zdot\ups64$ & $-28\arcd42\arcm13\zdot\arcs3$ & 16.96 & 1.36 & 0.417923 & $-7.21\pm0.55$~E$-7$ \\
241306 & $18\uph00\upm17\zdot\ups74$ & $-29\arcd16\arcm25\zdot\arcs1$ & 14.38 & 1.09 & 0.371268 & $-7.11\pm0.22$~E$-7$ \\
246693 & $18\uph00\upm46\zdot\ups32$ & $-28\arcd52\arcm27\zdot\arcs3$ & 15.69 & 0.84 & 0.363524 & $-7.01\pm0.32$~E$-7$ \\
122558 & $17\uph49\upm35\zdot\ups14$ & $-30\arcd48\arcm09\zdot\arcs5$ & 16.79 & 1.92 & 0.410946 & $-6.79\pm1.20$~E$-7$ \\
162594 & $17\uph53\upm33\zdot\ups56$ & $-30\arcd03\arcm08\zdot\arcs9$ & 16.79 & 1.70 & 0.356191 & $-6.24\pm0.26$~E$-7$ \\
207879 & $17\uph57\upm22\zdot\ups78$ & $-28\arcd56\arcm59\zdot\arcs1$ & 15.10 & 1.48 & 0.393082 & $-5.51\pm0.33$~E$-7$ \\
215121 & $17\uph58\upm00\zdot\ups28$ & $-29\arcd57\arcm49\zdot\arcs2$ & 17.46 & 1.59 & 0.307842 & $-5.51\pm1.13$~E$-7$ \\
250731 & $18\uph01\upm07\zdot\ups03$ & $-30\arcd42\arcm13\zdot\arcs7$ & 16.63 & 1.69 & 0.260272 & $-4.91\pm0.57$~E$-7$ \\
294795 & $18\uph05\upm10\zdot\ups60$ & $-29\arcd21\arcm03\zdot\arcs9$ & 13.04 & 1.02 & 0.293658 & $-4.84\pm0.41$~E$-7$ \\
198303 & $17\uph56\upm33\zdot\ups75$ & $-30\arcd14\arcm33\zdot\arcs8$ & 14.81 & 1.20 & 0.438734 & $-3.81\pm0.72$~E$-7$ \\
168012 & $17\uph53\upm59\zdot\ups73$ & $-30\arcd04\arcm48\zdot\arcs6$ & 16.40 & 1.58 & 0.371039 & $-3.81\pm0.68$~E$-7$ \\
279326 & $18\uph03\upm46\zdot\ups01$ & $-29\arcd47\arcm40\zdot\arcs6$ & 15.77 & 1.52 & 0.227733 & $-3.29\pm0.24$~E$-7$ \\
187430 & $17\uph55\upm38\zdot\ups54$ & $-29\arcd17\arcm24\zdot\arcs1$ & 17.47 & 2.01 & 0.275312 & $-3.18\pm0.54$~E$-7$ \\
217596 & $17\uph58\upm12\zdot\ups70$ & $-28\arcd48\arcm16\zdot\arcs6$ & 15.91 & 1.35 & 0.354705 & $-3.06\pm0.13$~E$-7$ \\
\hline
\noalign{\vskip3pt}
\end{tabular}}
\end{table}

\begin{figure}[htb]
\centerline{\includegraphics[angle=0,width=130mm]{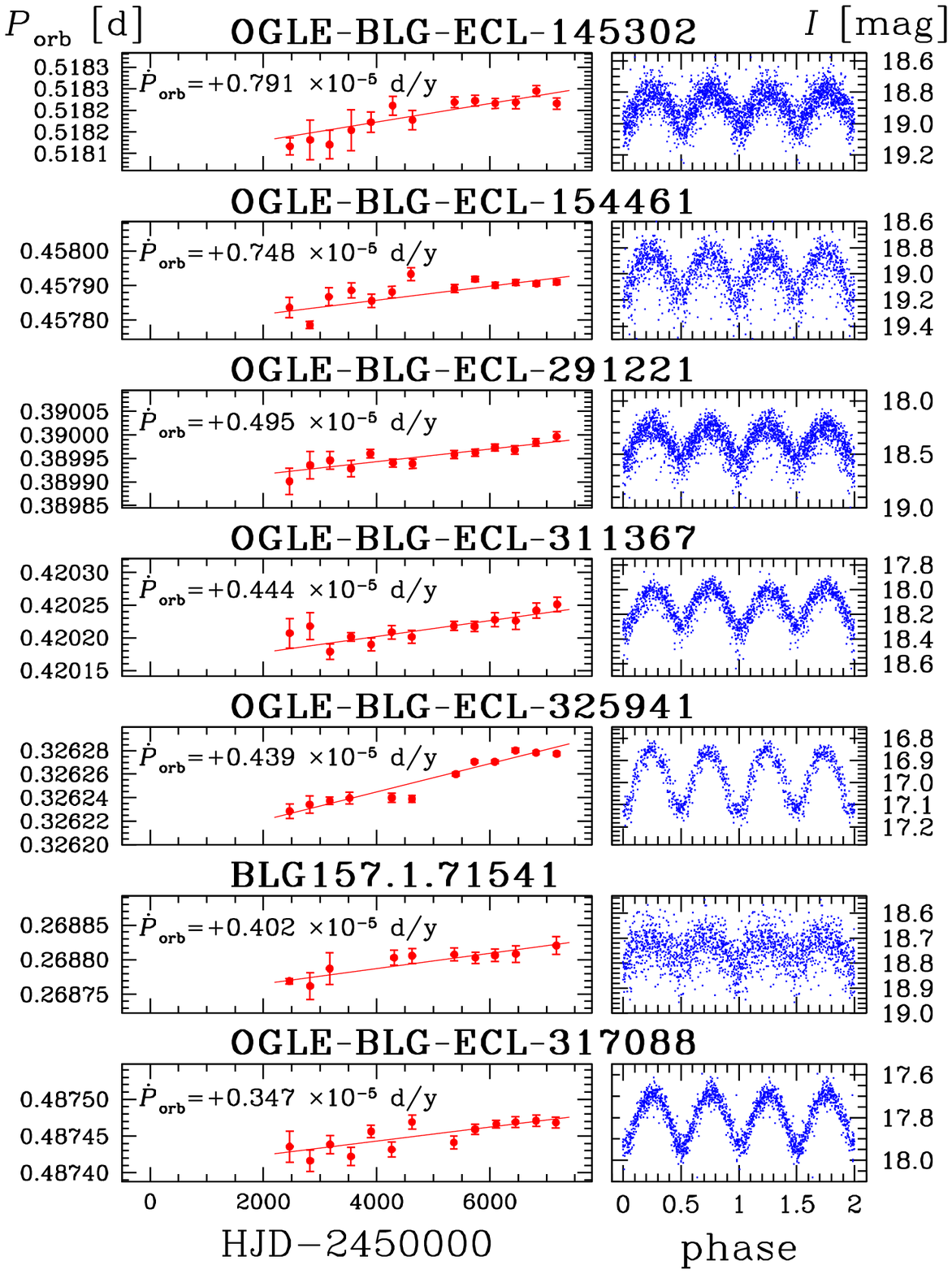}}
\FigCap{Binary systems with the highest positive period change rates.
OGLE-III light curves are presented.}
\end{figure}

\begin{table}[h!]
\centering
\caption{\small Parameters of the systems with derived positive period change rate}
\medskip
{\footnotesize
\begin{tabular}{ccccccc}
\hline
OGLE-BLG-ECL- &   RA    &    Dec  & $I_{\rm max}$ & $V-I$ & $P_{\rm orb}$ & $\dot P_{\rm orb}$ \\
              & J2000.0 & J2000.0 &     [mag]     & [mag] &      [d]      &         [d/y] \\
\hline
145302 & $17\uph52\upm04\zdot\ups81$ & $-30\arcd35\arcm24\zdot\arcs0$ & 18.74 & 2.89 & 0.518252 & $7.91\pm1.32$~E$-6$ \\
154461 & $17\uph52\upm52\zdot\ups57$ & $-30\arcd57\arcm27\zdot\arcs1$ & 18.80 & 1.84 & 0.457895 & $7.48\pm0.82$~E$-6$ \\
291221 & $18\uph04\upm50\zdot\ups83$ & $-29\arcd33\arcm26\zdot\arcs4$ & 18.16 & 1.92 & 0.389955 & $4.95\pm0.85$~E$-6$ \\
311367 & $18\uph06\upm47\zdot\ups01$ & $-30\arcd47\arcm49\zdot\arcs2$ & 17.94 & 1.44 & 0.420201 & $4.44\pm0.80$~E$-6$ \\
325941 & $18\uph08\upm19\zdot\ups42$ & $-26\arcd00\arcm05\zdot\arcs2$ & 16.84 & 1.62 & 0.326240 & $4.39\pm0.21$~E$-6$ \\
BLG157.1.71541  & $17\uph58\upm30\zdot\ups84$ & $-32\arcd39\arcm53\zdot\arcs8$ & 18.68 & 1.91 & 0.268803 & $4.02\pm0.59$~E$-6$ \\
317088 & $18\uph07\upm22\zdot\ups46$ & $-29\arcd18\arcm39\zdot\arcs2$ & 17.68 & 1.33 & 0.487443 & $3.47\pm0.69$~E$-6$ \\
183497 & $17\uph55\upm19\zdot\ups37$ & $-32\arcd55\arcm41\zdot\arcs0$ & 17.77 & 1.18 & 0.523197 & $3.36\pm0.79$~E$-6$ \\
126797 & $17\uph50\upm07\zdot\ups17$ & $-30\arcd09\arcm38\zdot\arcs1$ & 17.75 & 2.56 & 0.367062 & $3.36\pm0.57$~E$-6$ \\
306303 & $18\uph06\upm16\zdot\ups05$ & $-29\arcd00\arcm46\zdot\arcs8$ & 17.88 & 1.15 & 0.415518 & $3.31\pm0.59$~E$-6$ \\
243435 & $18\uph00\upm28\zdot\ups54$ & $-30\arcd27\arcm58\zdot\arcs7$ & 18.31 & 1.56 & 0.419818 & $3.25\pm0.58$~E$-6$ \\
201168 & $17\uph56\upm48\zdot\ups09$ & $-29\arcd30\arcm10\zdot\arcs7$ & 18.16 & 1.92 & 0.465644 & $2.90\pm0.56$~E$-6$ \\
200942 & $17\uph56\upm46\zdot\ups93$ & $-30\arcd00\arcm57\zdot\arcs6$ & 18.30 & 2.08 & 0.358376 & $2.86\pm0.52$~E$-6$ \\
122590 & $17\uph49\upm35\zdot\ups42$ & $-30\arcd32\arcm44\zdot\arcs1$ & 16.33 & 1.48 & 0.306970 & $2.36\pm0.10$~E$-6$ \\
169859 & $17\uph54\upm08\zdot\ups29$ & $-29\arcd44\arcm38\zdot\arcs1$ & 18.13 & 1.50 & 0.393950 & $2.34\pm0.13$~E$-6$ \\
245597 & $18\uph00\upm40\zdot\ups49$ & $-28\arcd39\arcm52\zdot\arcs1$ & 17.99 & 1.50 & 0.402376 & $2.29\pm0.11$~E$-6$ \\
197927 & $17\uph56\upm32\zdot\ups09$ & $-29\arcd15\arcm43\zdot\arcs0$ & 17.22 & 1.76 & 0.545565 & $2.26\pm0.27$~E$-6$ \\
203756 & $17\uph57\upm00\zdot\ups79$ & $-29\arcd39\arcm38\zdot\arcs4$ & 17.95 & 1.71 & 0.418336 & $2.14\pm0.41$~E$-6$ \\
192324 & $17\uph56\upm03\zdot\ups39$ & $-29\arcd25\arcm10\zdot\arcs6$ & 17.81 & 2.02 & 0.436902 & $2.05\pm0.24$~E$-6$ \\
282055 & $18\uph04\upm00\zdot\ups73$ & $-28\arcd39\arcm33\zdot\arcs7$ & 17.61 & 1.39 & 0.482814 & $2.04\pm0.14$~E$-6$ \\
BLG206.4.258074 & $18\uph00\upm56\zdot\ups13$ & $-28\arcd35\arcm56\zdot\arcs0$ & 18.07 & 1.66 & 0.405867 & $2.00\pm0.22$~E$-6$ \\
216089 & $17\uph58\upm05\zdot\ups00$ & $-28\arcd38\arcm34\zdot\arcs9$ & 15.77 & 1.48 & 0.490068 & $1.92\pm0.12$~E$-6$ \\
196868 & $17\uph56\upm26\zdot\ups44$ & $-29\arcd18\arcm34\zdot\arcs7$ & 16.12 & 1.70 & 0.400507 & $1.86\pm0.09$~E$-6$ \\
233754 & $17\uph59\upm37\zdot\ups55$ & $-29\arcd12\arcm27\zdot\arcs8$ & 16.86 & 1.35 & 0.570240 & $1.84\pm0.09$~E$-6$ \\
157410 & $17\uph53\upm07\zdot\ups74$ & $-30\arcd27\arcm52\zdot\arcs4$ & 15.39 & 1.35 & 0.363414 & $1.69\pm0.08$~E$-6$ \\
124567 & $17\uph49\upm50\zdot\ups90$ & $-29\arcd38\arcm17\zdot\arcs9$ & 16.20 & 1.78 & 0.400343 & $1.61\pm0.03$~E$-6$ \\
233878 & $17\uph59\upm38\zdot\ups27$ & $-28\arcd45\arcm47\zdot\arcs8$ & 13.80 & 0.92 & 0.741434 & $1.60\pm0.04$~E$-6$ \\
194517 & $17\uph56\upm14\zdot\ups16$ & $-29\arcd43\arcm37\zdot\arcs7$ & 17.20 & 1.57 & 0.421592 & $1.58\pm0.20$~E$-6$ \\
340259 & $18\uph09\upm55\zdot\ups74$ & $-29\arcd33\arcm39\zdot\arcs6$ & 14.77 & 0.79 & 0.453261 & $1.38\pm0.19$~E$-6$ \\
199259 & $17\uph56\upm38\zdot\ups43$ & $-29\arcd48\arcm45\zdot\arcs8$ & 13.74 & 0.97 & 0.389284 & $1.37\pm0.09$~E$-6$ \\
270325 & $18\uph02\upm56\zdot\ups52$ & $-30\arcd17\arcm39\zdot\arcs6$ & 16.59 & 1.38 & 0.381891 & $1.32\pm0.11$~E$-6$ \\
230641 & $17\uph59\upm20\zdot\ups86$ & $-28\arcd46\arcm18\zdot\arcs7$ & 15.56 & 1.24 & 0.382904 & $1.29\pm0.03$~E$-6$ \\
195044 & $17\uph56\upm16\zdot\ups92$ & $-30\arcd44\arcm39\zdot\arcs6$ & 16.39 & 1.47 & 0.409737 & $1.16\pm0.03$~E$-6$ \\
135055 & $17\uph51\upm04\zdot\ups99$ & $-29\arcd40\arcm10\zdot\arcs7$ & 15.75 & 1.69 & 0.522241 & $1.10\pm0.13$~E$-6$ \\
153698 & $17\uph52\upm48\zdot\ups61$ & $-29\arcd06\arcm42\zdot\arcs9$ & 16.72 & 1.69 & 0.386626 & $1.08\pm0.14$~E$-6$ \\
234160 & $17\uph59\upm39\zdot\ups66$ & $-29\arcd05\arcm13\zdot\arcs9$ & 16.94 & 1.35 & 0.528657 & $1.06\pm0.06$~E$-6$ \\
200313 & $17\uph56\upm43\zdot\ups77$ & $-30\arcd49\arcm34\zdot\arcs2$ & 17.24 & 1.54 & 0.479131 & $1.03\pm0.05$~E$-6$ \\
174839 & $17\uph54\upm34\zdot\ups06$ & $-29\arcd24\arcm38\zdot\arcs3$ & 16.42 & 1.46 & 0.461225 & $9.62\pm0.24$~E$-7$ \\
219977 & $17\uph58\upm24\zdot\ups47$ & $-28\arcd41\arcm05\zdot\arcs4$ & 15.62 & 1.65 & 0.486807 & $8.94\pm0.31$~E$-7$ \\
127233 & $17\uph50\upm10\zdot\ups38$ & $-35\arcd02\arcm26\zdot\arcs6$ & 16.73 & 1.44 & 0.350237 & $8.74\pm1.59$~E$-7$ \\
205743 & $17\uph57\upm12\zdot\ups10$ & $-29\arcd56\arcm21\zdot\arcs1$ & 15.75 & 1.40 & 0.300489 & $8.24\pm0.97$~E$-7$ \\
212539 & $17\uph57\upm45\zdot\ups97$ & $-29\arcd12\arcm04\zdot\arcs9$ & 17.61 & 1.79 & 0.286102 & $8.15\pm1.49$~E$-7$ \\
317072 & $18\uph07\upm22\zdot\ups35$ & $-29\arcd42\arcm13\zdot\arcs1$ & 15.53 & 1.25 & 0.353712 & $7.28\pm0.82$~E$-7$ \\
162146 & $17\uph53\upm31\zdot\ups37$ & $-29\arcd53\arcm13\zdot\arcs0$ & 17.13 & 1.87 & 0.410687 & $5.97\pm0.62$~E$-7$ \\
125481 & $17\uph49\upm57\zdot\ups46$ & $-29\arcd16\arcm03\zdot\arcs0$ & 15.08 & 1.60 & 0.358558 & $5.88\pm0.27$~E$-7$ \\
207420 & $17\uph57\upm20\zdot\ups44$ & $-30\arcd36\arcm58\zdot\arcs6$ & 14.27 & 1.02 & 0.254388 & $5.63\pm0.11$~E$-7$ \\
158555 & $17\uph53\upm13\zdot\ups46$ & $-31\arcd13\arcm56\zdot\arcs5$ & 17.60 & 1.96 & 0.231003 & $5.44\pm0.48$~E$-7$ \\
284531 & $18\uph04\upm14\zdot\ups75$ & $-29\arcd52\arcm19\zdot\arcs8$ & 13.56 & 0.93 & 0.435555 & $5.17\pm0.21$~E$-7$ \\
159557 & $17\uph53\upm18\zdot\ups25$ & $-29\arcd53\arcm13\zdot\arcs3$ & 14.37 & 0.98 & 0.522287 & $5.01\pm0.27$~E$-7$ \\
202706 & $17\uph56\upm55\zdot\ups80$ & $-28\arcd39\arcm16\zdot\arcs6$ & 15.88 & 1.68 & 0.449735 & $4.08\pm0.71$~E$-7$ \\
182350 & $17\uph55\upm13\zdot\ups65$ & $-29\arcd26\arcm44\zdot\arcs8$ & 15.55 & 1.29 & 0.282379 & $3.46\pm0.16$~E$-7$ \\
181955 & $17\uph55\upm11\zdot\ups74$ & $-30\arcd02\arcm12\zdot\arcs1$ & 16.14 & 1.41 & 0.414302 & $2.62\pm0.14$~E$-7$ \\
\hline
\noalign{\vskip3pt}
\end{tabular}}
\end{table}

\begin{figure}[htb]
\centerline{\includegraphics[angle=0,width=130mm]{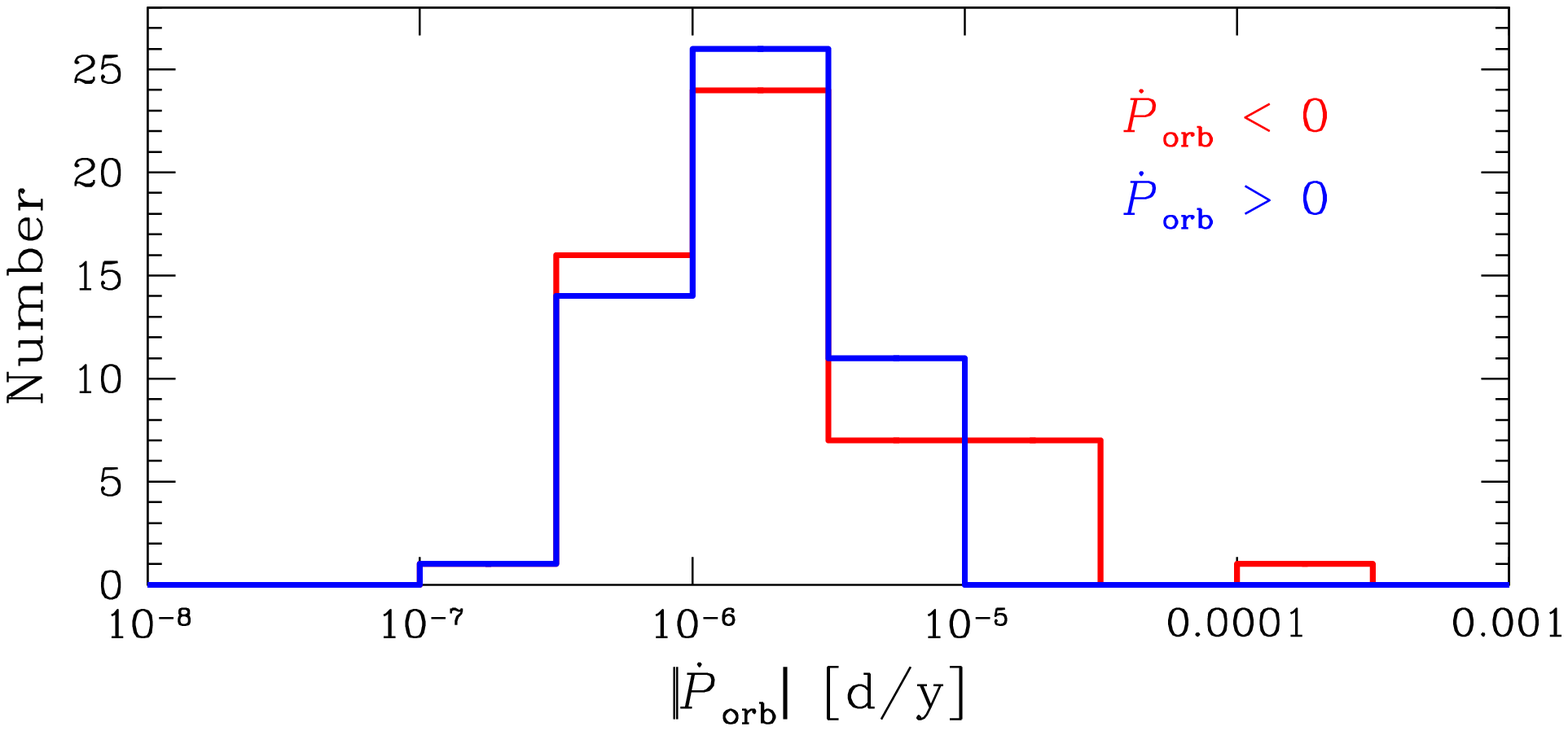}}
\FigCap{Comparison of the distributions of systems with negative and positive
period change rates. Note the excess of systems with high negative rate.}
\end{figure}

\begin{figure}[htb]
\centerline{\includegraphics[angle=0,width=130mm]{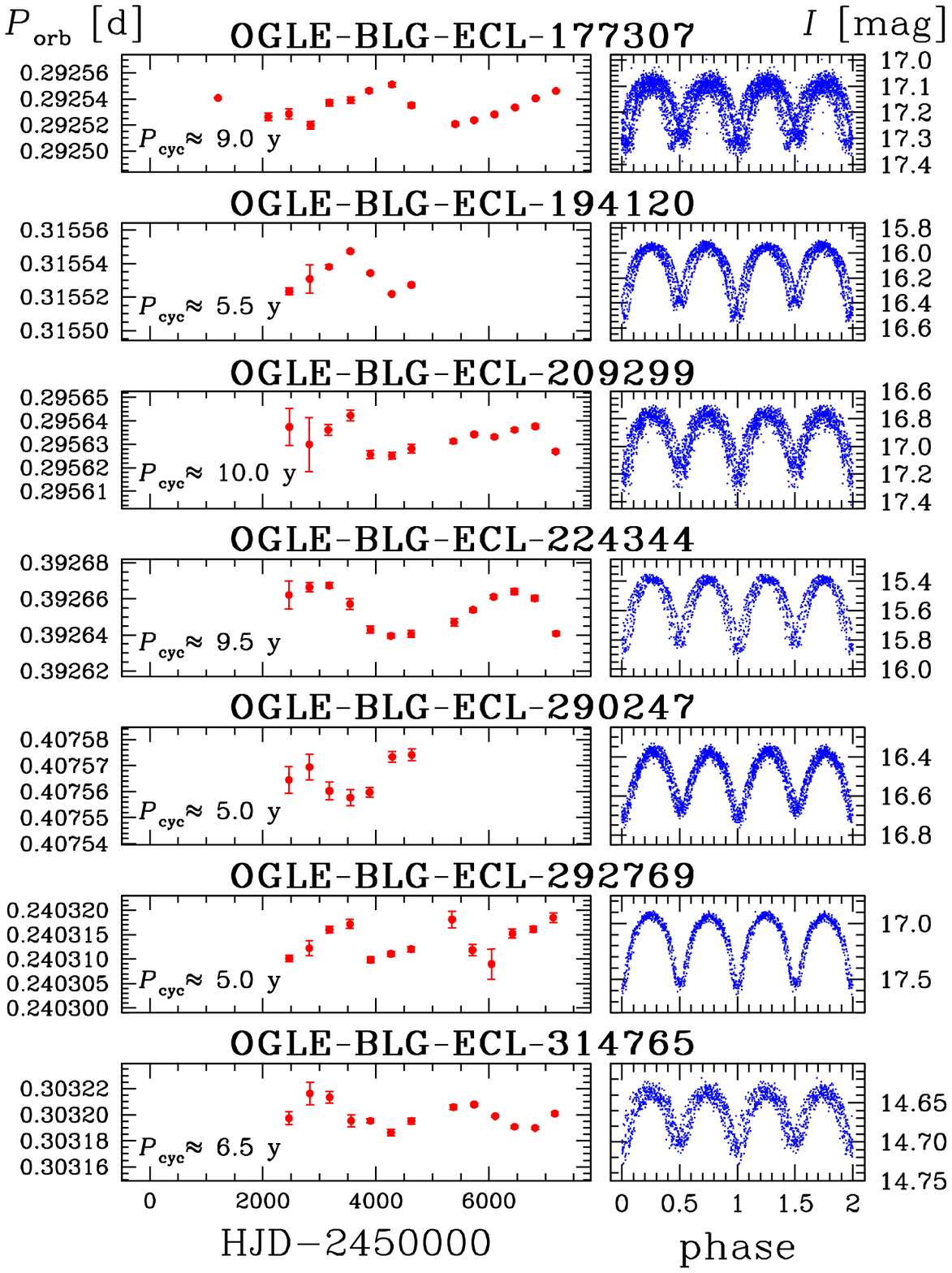}}
\FigCap{Example binary systems with cyclic orbital period variations.
OGLE-III light curves are phased with constant period determined
for this dataset.}
\end{figure}

\begin{table}[h!]
\centering
\caption{\small Parameters of candidate systems with cyclic period variations}
\medskip
{\footnotesize
\begin{tabular}{ccccccc}
\hline
OGLE-BLG-ECL- &   RA    &    Dec  & $I_{\rm max}$ & $V-I$ & $P_{\rm orb}$ & $P_{\rm cyc}$ \\
              & J2000.0 & J2000.0 &     [mag]     & [mag] &      [d]      &     [y] \\
\hline
124213 & $17\uph49\upm48\zdot\ups43$ & $-33\arcd44\arcm49\zdot\arcs2$ & 18.34 & 1.96 & 0.83464370 &  6   \\
124728 & $17\uph49\upm51\zdot\ups95$ & $-29\arcd41\arcm48\zdot\arcs8$ & 17.35 & 2.10 & 0.35675024 &  6.5 \\
125525 & $17\uph49\upm57\zdot\ups66$ & $-34\arcd39\arcm54\zdot\arcs5$ & 14.76 & 1.20 & 0.41200122 & 15   \\
127744 & $17\uph50\upm14\zdot\ups07$ & $-30\arcd06\arcm22\zdot\arcs3$ & 14.98 & 1.77 & 0.89206984 &  6.5 \\
129787 & $17\uph50\upm29\zdot\ups23$ & $-29\arcd42\arcm24\zdot\arcs8$ & 16.43 & 2.11 & 0.60583946 &  7   \\
144092 & $17\uph51\upm58\zdot\ups17$ & $-29\arcd46\arcm41\zdot\arcs5$ & 15.54 & 1.58 & 0.37023316 &  5   \\
148291 & $17\uph52\upm21\zdot\ups01$ & $-29\arcd46\arcm59\zdot\arcs3$ & 16.94 & 1.59 & 0.33885024 & 21   \\
153387 & $17\uph52\upm47\zdot\ups02$ & $-33\arcd14\arcm16\zdot\arcs0$ & 18.09 & 1.78 & 0.41679258 & 14   \\
154199 & $17\uph52\upm51\zdot\ups23$ & $-29\arcd46\arcm23\zdot\arcs6$ & 16.29 & 1.50 & 0.35765448 &  9   \\
157195 & $17\uph53\upm06\zdot\ups58$ & $-30\arcd06\arcm54\zdot\arcs9$ & 16.25 & 1.59 & 0.30197516 &  3.0 \\
163206 & $17\uph53\upm36\zdot\ups89$ & $-32\arcd43\arcm33\zdot\arcs6$ & 17.65 &  $-$ & 0.32505982 & 16   \\
163661 & $17\uph53\upm38\zdot\ups87$ & $-29\arcd39\arcm03\zdot\arcs8$ & 18.56 & 1.81 & 0.30801484 &  6.5 \\
163907 & $17\uph53\upm40\zdot\ups05$ & $-30\arcd07\arcm16\zdot\arcs9$ & 15.25 & 1.37 & 0.28606972 &  7   \\
168647 & $17\uph54\upm02\zdot\ups50$ & $-33\arcd00\arcm18\zdot\arcs2$ & 16.10 & 1.65 & 0.31771956 & 10   \\
171861 & $17\uph54\upm18\zdot\ups15$ & $-29\arcd33\arcm30\zdot\arcs8$ & 17.62 & 2.03 & 0.33601454 &  2.0 \\
174546 & $17\uph54\upm32\zdot\ups42$ & $-29\arcd34\arcm58\zdot\arcs1$ & 15.66 & 1.31 & 0.45498264 & 20   \\
177307 & $17\uph54\upm46\zdot\ups24$ & $-29\arcd44\arcm55\zdot\arcs2$ & 17.08 & 1.64 & 0.29253374 &  9   \\
184596 & $17\uph55\upm24\zdot\ups33$ & $-29\arcd33\arcm41\zdot\arcs8$ & 13.57 & 0.98 & 0.52178366 &  1.5 \\
192607 & $17\uph56\upm04\zdot\ups85$ & $-30\arcd20\arcm50\zdot\arcs3$ & 18.06 & 2.46 & 0.36283864 &  4   \\
194120 & $17\uph56\upm12\zdot\ups29$ & $-30\arcd46\arcm02\zdot\arcs8$ & 15.93 & 1.47 & 0.31553406 &  5.5 \\
207113 & $17\uph57\upm19\zdot\ups01$ & $-33\arcd53\arcm47\zdot\arcs1$ & 17.23 & 1.13 & 0.44961886 &  2.7 \\
208059 & $17\uph57\upm23\zdot\ups64$ & $-30\arcd49\arcm15\zdot\arcs3$ & 16.52 & 1.53 & 0.41324802 &  1.5 \\
209299 & $17\uph57\upm29\zdot\ups74$ & $-30\arcd53\arcm29\zdot\arcs3$ & 16.76 & 1.72 & 0.29563266 & 10   \\
219570 & $17\uph58\upm22\zdot\ups46$ & $-29\arcd45\arcm49\zdot\arcs1$ & 16.48 & 1.39 & 0.31779704 &  5   \\
222038 & $17\uph58\upm35\zdot\ups09$ & $-29\arcd53\arcm02\zdot\arcs3$ & 18.38 & 1.77 & 0.34785152 &  4.5 \\
224344 & $17\uph58\upm47\zdot\ups91$ & $-26\arcd50\arcm46\zdot\arcs5$ & 15.38 & 1.52 & 0.39265194 &  9.5 \\
235487 & $17\uph59\upm46\zdot\ups18$ & $-29\arcd16\arcm00\zdot\arcs4$ & 15.90 & 1.35 & 0.46023396 &  6.5 \\
240451 & $18\uph00\upm13\zdot\ups09$ & $-29\arcd14\arcm25\zdot\arcs9$ & 16.70 & 1.39 & 0.36804430 & 22   \\
246852 & $18\uph00\upm47\zdot\ups15$ & $-28\arcd36\arcm58\zdot\arcs2$ & 17.57 & 1.55 & 0.42095148 &  8.5 \\
250500 & $18\uph01\upm05\zdot\ups72$ & $-30\arcd13\arcm22\zdot\arcs7$ & 18.27 & 1.71 & 0.47156330 &  7   \\
289401 & $18\uph04\upm41\zdot\ups02$ & $-28\arcd46\arcm03\zdot\arcs7$ & 16.91 & 1.22 & 0.48576687 &  2.8 \\
290247 & $18\uph04\upm45\zdot\ups58$ & $-29\arcd31\arcm37\zdot\arcs7$ & 16.36 & 1.47 & 0.40756464 &  5   \\
292769 & $18\uph04\upm59\zdot\ups46$ & $-30\arcd36\arcm47\zdot\arcs3$ & 16.92 & 1.86 & 0.24031322 &  5   \\
313336 & $18\uph06\upm58\zdot\ups97$ & $-27\arcd41\arcm34\zdot\arcs2$ & 18.70 & 1.60 & 0.32255660 &  3.5 \\
314765 & $18\uph07\upm07\zdot\ups94$ & $-29\arcd39\arcm00\zdot\arcs9$ & 14.63 & 1.02 & 0.30319922 &  6.5 \\
\hline
\noalign{\vskip3pt}
\end{tabular}}
\end{table}


\Section{Summary and Conclusions}

In the huge sample of 22~462 eclipsing binaries with $P_{\rm orb}<4$~d
detected in the OGLE-III Galactic bulge fields, we found 108 systems
with reliable monotonic period changes: 56 systems with negative
rate and 52 systems with positive one. We also indicated 35 systems with
evident cyclic period variations. All reported systems but object
OGLE-BLG-ECL-139622 with the highest derived negative period change rate
are contact binaries. We did not find any binary with rapid
orbital period decrease as expected for a system heading toward merger
within a few years. The period change rate in V1309 Sco five years
before the merger event was about $-8.3\times10^{-4}$~d/y,
while two years before the event it reached $-3.8\times10^{-3}$~d/y.
For comparison, contact binary OGLE-BLG-ECL-299145 with the second
fastest measured period decrease in the whole our sample has the
rate of merely $-1.7\times10^{-5}$~d/y. Twenty-two our systems
with negative period changes have the absolute rate higher then
$-2\times10^{-6}$~d/y, that is the value estimated for contact system
KIC 9832227 by Molnar \etal (2017). We cannot exclude the possibility
that this particular binary system and also our contact binaries with
relatively long orbital period ($P_{\rm orb}>1.0$~d) and relatively
high negative period change rate ($|\dot P_{\rm orb}|>10^{-5}$~d/y), such
as OGLE-BLG-ECL-344477, OGLE-BLG-ECL-176377, and OGLE-BLG-ECL-170070,
will merge in near future (in tens or hundreds of years).
However, the fact that eclipsing binary OGLE-BLG-ECL-139622 with the
highest derived period change rate in our sample is a detached
system and all the remaining binaries with reliable period changes
are short-period contact binaries, some of which show cyclic variations,
strongly indicate for the presence of third bodies in the investigated
systems. Results from various observations of nearby close
binaries ($P_{\rm orb}<1.0$~d) support the idea that tertiary companions
to such binaries are very common (D'Angelo \etal 2006,
Pribulla and Rucinski 2006, Tokovinin \etal 2006, Rucinski \etal 2007).

Another possible explanation of the observed period changes could
be slow movement of starspots on the surface of the binary components.
Close binary systems are often chromospherically active and thus they
can be strong X-ray emitters. We looked for X-ray counterparts to our 143
eclipsing binaries with the detected period changes and we found that
all 32 binaries located within the Chandra Galactic Bulge Survey area
($-3\arcd \lesssim l \lesssim 3\arcd$, $1\arcd \lesssim |b| \lesssim 2\arcd$,
Jonker \etal 2011, Wevers \etal 2016) have such counterparts. Four
other binaries have counterparts in the XMM-Newton data (Page \etal 2012).
Slowly drifting starspots would result in long-term brightness variations
in the light curves of binaries. Some of our objects exhibit
such mean brightness variations (see examples in Fig.~6).

\begin{figure}[htb]
\centerline{\includegraphics[angle=0,width=130mm]{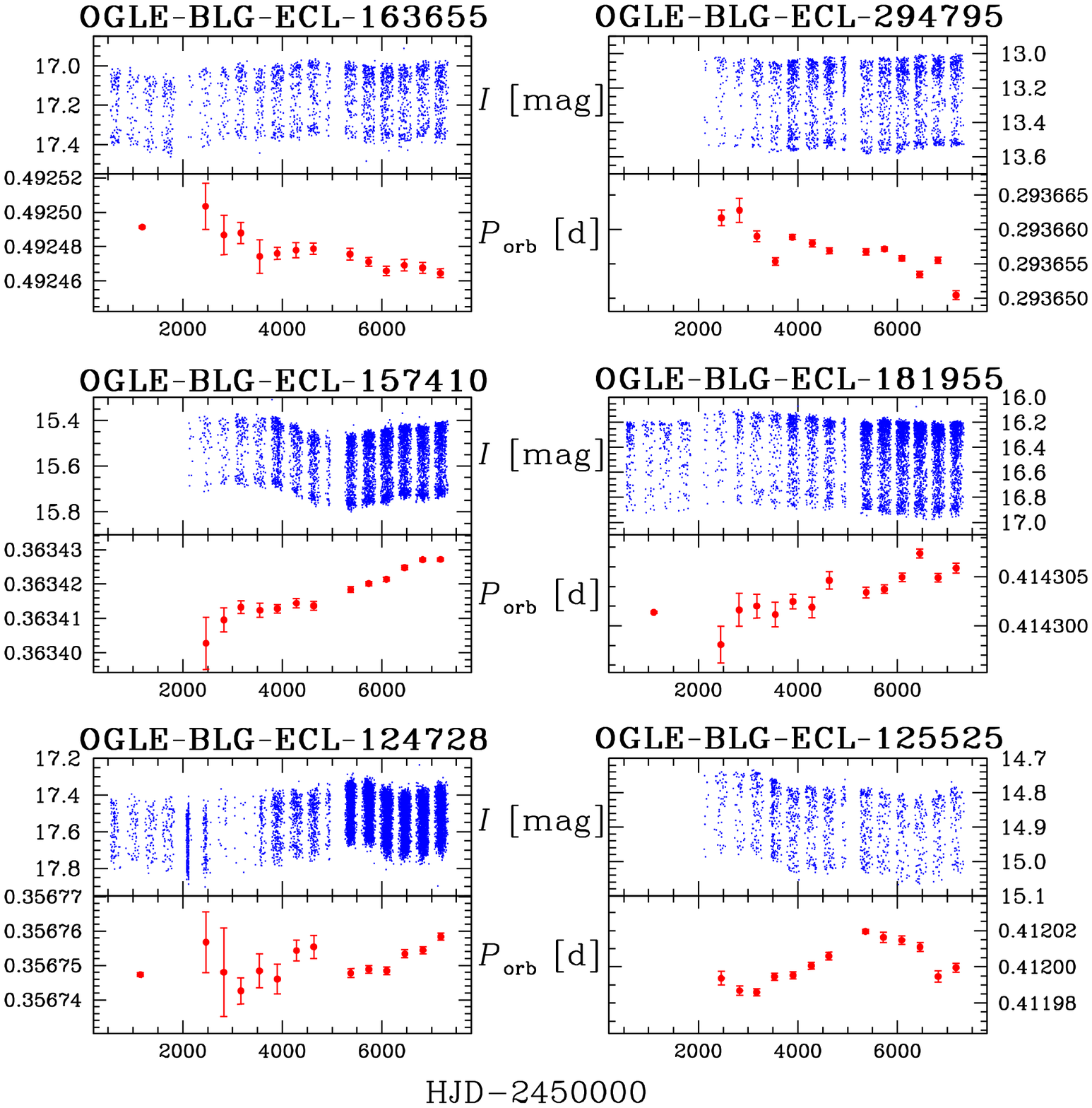}}
\FigCap{Long-term brightness variations in selected binary systems with
negative (upper panels), positive (middle panels), and cyclic period changes
(lower panels). In some cases, the brightness variations seem to correlate
with the period changes. All stars are isolated objects in the OGLE images.}
\end{figure}

The negative result of our search for contact systems with rapid period
decrease in the OGLE-III data is in agreement with the lack of observed
Galactic red nova outbursts during the current fourth phase of OGLE.
Despite our unsuccessful search for future mergers, binaries with relatively
long orbital period and high negative period change rate are worth further
monitoring.


\Acknow{
We would like to thank Profs. M. Kubiak and G. Pietrzy\'nski,
former members of the OGLE team, for their contribution to the
collection of the OGLE photometric data over the past years.
The OGLE project has received funding from the National Science
Centre, Poland, grant MAESTRO 2014/14/A/ST9/00121 to A.U. This work
has been also supported by the Polish Ministry of Sciences and Higher
Education grants No. IP2012 005672 under the Iuventus Plus program
to P.P. and No. IdP2012 000162 under the Ideas Plus program to I.S.}



\begin{references}

\refitem{Alard, C., and Lupton, R.H.}{1998}{\ApJ}{503}{325}
\refitem{Brown, N.J., Waagen, E.O., Scovil, C., Nelson, P., Oksanen, A., Solonen, J., and Price, A.}{2002}{IAU Circ.}{7785}{1}
\refitem{Conroy, K.E., Pr\v sa, A., Stassun, K.G., Orosz, J.A., Fabrycky, D.C., and Welsh, W.F.}{2014}{\AJ}{147}{45}
\refitem{D'Angelo, C., van Kerkwijk, M.H., and Rucinski, S.M.}{2006}{\AJ}{132}{650}
\refitem{Gies, D.R., Matson, R.A., Guo, Z., Lester, K.V., Orosz, J.A., and Peters, G.J.}{2015}{\AJ}{150}{178}
\refitem{Hayashi, S.S., Yamamoto, M., and Hirosawa, K.}{1994}{IAU Circ}{5942}{1}
\refitem{Jonker, P.G., \etal}{2011}{\ApJS}{194}{18}
\refitem{Kochanek, C.S., Adams, S.M., and Belczynski, K.}{2014}{\MNRAS}{443}{1319}
\refitem{Kubiak, M., Udalski, A., and Szyma\'nski, M.K.}{2006}{\Acta}{56}{253}
\refitem{Molnar, L.A. \etal}{2017}{\ApJ}{840}{1}
\refitem{Mr\'oz, P., \etal}{2015}{\Acta}{65}{313}
\refitem{Nakano, S., Nishiyama, K., Kabashima, F., and Sakurai, Y.}{2008}{Central Bureau Electronic Telegrams}{1496}{1}
\refitem{Page, M. J., \etal}{2012}{\MNRAS}{426}{903}
\refitem{Pejcha, O.}{2014}{\ApJ}{788}{22}
\refitem{Pietrukowicz, P., \etal}{2013}{\Acta}{63}{115}
\refitem{Pribulla, T., and Rucinski, S.M.}{2006}{\AJ}{131}{2986}
\refitem{Rucinski, S.M., Pribulla, T., and van Kerkwijk, M.H.}{2007}{\AJ}{134}{2353}
\refitem{Schwarzenberg-Czerny, A.}{1996}{\ApJ}{460}{L107}
\refitem{Shara, M.M., Yaron, O., Prialnik, D., Kovetz, A., and Zurek, D.}{2010}{\ApJ}{725}{831}
\refitem{Soker, N., and Tylenda, R.}{2003}{\ApJ}{582}{L105}
\refitem{Soszy\'nski, I., \etal}{2011}{\Acta}{61}{1}
\refitem{Soszy\'nski, I., \etal}{2013}{\Acta}{63}{21}
\refitem{Soszy\'nski, I., \etal}{2014}{\Acta}{64}{177}
\refitem{Soszy\'nski, I., \etal}{2015}{\Acta}{65}{39}
\refitem{Soszy\'nski, I., \etal}{2016}{\Acta}{66}{405}
\refitem{Szyma\'nski, M.K., \etal}{2011}{\Acta}{61}{83}
\refitem{Thompson, T.A., Prieto, J.L., Stanek, K.Z., Kistler, M.D., Beacom, J.F., and Kochanek, C.S.}{2009}{\ApJ}{705}{1364}
\refitem{Tokovinin, A., Thomas, S., Sterzik, M., and Udry, S.}{2006}{\AA}{450}{681}
\refitem{Tylenda, R., \etal}{2011}{\AA}{528A}{114}
\refitem{Tylenda, R., \etal}{2013}{\AA}{555}{A16}
\refitem{Udalski, A., Szyma\'nski, M., Ka{\l}u\.zny, J., Kubiak, M., and Mateo, M.}{1992}{\Acta}{42}{253}
\refitem{Udalski, A., Kubiak, M., and Szyma\'nski, M.K.}{1997}{\Acta}{47}{169}
\refitem{Udalski, A.}{2003}{\Acta}{53}{291}
\refitem{Udalski, A., Szyma\'nski, M.K., and Szyma\'nski, G.}{2015}{AA}{65}{1}
\refitem{Wevers, T., \etal}{2016}{\MNRAS}{458}{4530}
\refitem{Wo\'zniak, P.R.}{2000}{\Acta}{50}{421}

\end{references}
\end{document}